\documentclass[aps,prl,10pt,showpacs,superscriptaddress,twocolumn,amssymb,amsfonts]{revtex4-1}

\usepackage{amsmath,bm}
\usepackage{graphics}
\usepackage{dsfont}
\usepackage{subfigure}		
\usepackage{epsfig}
\usepackage{hyperref}		
\usepackage{units}

\newcommand{\xp}[1]{ {\left\langle{#1}\right\rangle} }

\newcommand{\phd}{ {\vphantom{\dag}} }		

\newcommand{\va}{ {\mathbf{a}} }
\newcommand{\vk}{ {\mathbf{k}} }
\newcommand{\vG}{ {\mathbf{G}} }

\renewcommand{\vr}{ {\mathbf{r}} }

\usepackage[usenames,dvipsnames]{color}			
\definecolor{purple}{rgb}{0.5,0,0.5}

\begin{document}
\title{Quantum transport and two-parameter scaling at the surface of a weak topological insulator}

\author{Roger S.~K.~Mong}
	\affiliation{Department of Physics, University of California, Berkeley, California 94720, USA}
\author{Jens H. Bardarson}
	\affiliation{Department of Physics, University of California, Berkeley, California 94720, USA}
	\affiliation{Materials Sciences Division, Lawrence Berkeley National Laboratory, Berkeley, California 94720}
\author{Joel E. Moore}
	\affiliation{Department of Physics, University of California, Berkeley, California 94720, USA}
	\affiliation{Materials Sciences Division, Lawrence Berkeley National Laboratory, Berkeley, California 94720}
\date{\today}

\begin{abstract}
Weak topological insulators have an even number of Dirac cones in their surface spectrum and are thought to be unstable to disorder, which leads to an insulating surface.
Here we argue that the presence of disorder alone will not localize the surface states,
rather; the presence of a time-reversal symmetric mass term is required for localization.
Through numerical simulations, we show that in the absence of the mass term the surface always flow to a stable metallic phase and the conductivity obeys a one-parameter scaling relation, just as in the case of a strong topological insulator surface.
With the inclusion of the mass, the transport properties of the surface of a weak topological insulator follow a two-parameter scaling form.
\end{abstract}


\maketitle


The conventional band theory categorizes crystals as metals, semiconductors, or insulators depending on the size of their band gap. Over the
last few years it has been realized that this categorization overlooks the fundamental fact that not all insulators are equivalent.  
Topological insulators are characterized by nontrivial band topology leading to gapless metallic surface states which are robust to disorder that preserves time-reversal symmetry (TRS)%
	~\cite{HasanKaneTIReview,*HasanMoore3DTIReview,*QiSCZhangTITSCReview}.
In two dimensions (2D), the quantum spin Hall (QSH) insulator possesses a pair of counterpropagating edge modes which are protected from backscattering by TRS.
In three dimensions (3D), topological insulators are classified as either strong (STI) or weak topological insulators (WTI).
The surfaces of STIs have an odd number of 2D Dirac fermions and have garnered much of the attention,
as TRS disorder cannot localize the surface states unless it is strong enough to move states across the bulk energy gap.
In contrast, the WTIs have an even number of Dirac fermions and are believed to be unstable to disorder%
	~\cite{HasanKaneTIReview,*HasanMoore3DTIReview,*QiSCZhangTITSCReview}.

This belief stems partially from comparisons with graphene.
Superficially, WTIs and graphene are similar in that their low energy electronic properties are described by an even number of Dirac fermions%
	~\cite{PeresGrapheneElectronicRev09,*DasSarmaGrapheneTransportRev11}.
While both systems have TRS implemented by an antiunitary time-reversal operator $\Theta$,
they differ fundamentally in that $\Theta^2 = +1$ for graphene from $\mathrm{SU}(2)$ spin symmetry%
	~\footnote{An unbroken $\mathrm{SU}(2)$ spin symmetry (\textit{e.g.}\ in graphene) means that $\Theta$ can be taken just to be complex conjugation, effectively treating the fermions as spinless with $\Theta^2 = +1$.},
	while for a WTI $\Theta^2 = -1$ due to the presence of strong spin-orbit coupling.
This places graphene in the orthogonal (AI) symmetry class, while WTIs belong to the symplectic (AII) class in the Altland-Zirnbauer classification%
	~\cite{AltlandZirnbauer97}.
The consequences of the minus sign are profound.
The first quantum correction to the Drude conductivity is determined by interference of time-reversal-symmetric paths.
In the orthogonal class, this interference is constructive (weak localization) and eventually leads to localization of all single particle states.
In contrast, in the symplectic class, the interference is destructive (weak antilocalization) giving rise to an enhancement of the conductivity and a stable \emph{symplectic metal}
phase~\cite{Hikami2DDisorder80}.
Hence, the metallic phase of graphene is unstable to disorder coupling the Dirac fermions%
	~\cite{AleinerEfetovGphDisorder06}
	but is stable in WTIs.

An STI is also in the symplectic class.
With an odd number of Dirac fermions on its surface, it always flows into the symplectic metal
	~\cite{BardarsonSingleParaScaling07,NomuraDiracDelocalization07},
reflecting the presence of a topological term in the effective field theory (nonlinear sigma model) describing diffusion%
	~\cite{OstrovskySymplecticTopTerm07,RyuSymplecticZ2TopTerm07}.
This topological term is absent in the same description of a WTI, suggesting that localization should occur.
In conventional semiconductors with spin-orbit coupling, this leads to a metal-insulator transition at a 
critical conductivity $\sigma_c \approx 1.42\,e^2/h$%
	~\cite{MarkosSchweitzerCriticalCond06}.

It is the purpose of this work to explore the precise conditions under which a WTI undergoes localization.
One reason that this is a pressing question is the following argument~\cite{RingelKrausSternWTI11}.
If one considers obtaining a WTI by stacking 2D layers in the QSH phase,
a surface parallel to the stacking direction would consist of pairs of one-dimensional (1D) counterpropagating helical modes.
The number of such modes taking part in transport can be even or odd depending on the number of layers.
However, an odd number of 1D modes in the symplectic class
necessarily leads to the presence of a perfectly transmitted mode
and thus a minimum conductance of $e^2/h$%
	~\cite{AndoSuzuuraCondChannel02,EversMirlinAndersonTransRev08}.
While this argument is one-dimensional in nature as the sample thickness is constant, it suggests that a WTI can under certain conditions avoid localization.
In the extended two-dimensional surface, the meaning of this parity effect is unclear, raising the question: What is the scaling behavior of the conductivity in disordered WTIs?

\begin{figure}[tb!]
	\includegraphics[width=0.6\columnwidth]{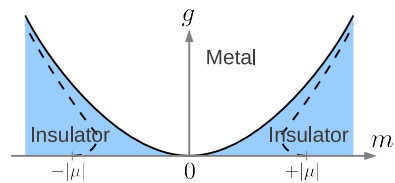}
	\caption{%
		The phase diagram of the Hamiltonian \eqref{eq:H} as a function of mass $m$ and disorder strength $g_{\alpha\beta} = g$.
		The solid line marks the metal-insulator transition at $\mu = 0$, whereas the dashed line marks the transition at finite $\mu$.
		At the clean Dirac point ($g = \mu=0$), there is a topological phase transition between the two types of insulators.
		With increasing disorder or chemical potential $\mu$, a metallic phase appears separating the two topological sectors.
	}
	\label{fig:QSHMetalIns}
\end{figure}

In this Letter, we demonstrate, by numerical simulations, that the scaling flow depends on the presence or absence of a specific time-reversal-symmetric mass, to be defined below.
In the presence of this mass, a gap opens up in the spectrum which can lead to localization. Disorder can still drive
the system into a metallic phase, realizing a metal-insulator transition at a critical value of conductivity consistent with what is observed in
conventional semiconductors. In contrast, in the absence of this mass the system always flows into the symplectic metal. We demonstrate that this flow follows one-parameter scaling with a positive beta function, just as in the case of an
STI~\cite{BardarsonSingleParaScaling07,NomuraDiracDelocalization07}.
The phase diagram emerging from these observations (\textit{cf.}\ Fig.~\ref{fig:QSHMetalIns})
suggests that one-parameter scaling is not realized throughout, as one might expect from the minimal nonlinear sigma model description.
Instead, we present data supporting two-parameter scaling, the effective field theory of which remains unknown.

\begin{table}[tb!]
\begin{tabular}{cc@{\quad}c}
		Disorder structure	&	Disorder type	&	Notation
\\	\hline
	$V_{x0} \cdot \tau^x$	&	scalar potential ($2\times$AII)	\\
	$V_{yx} \cdot \tau^y\sigma^x$	&	gauge potential ($2\times$AIII)	\\
	$V_{yy} \cdot \tau^y\sigma^y$	&	gauge potential ($2\times$AIII)	\\
	$V_{yz} \cdot \tau^y\sigma^z$	&	mass ($2\times$D)	&	$m = \xp{V_{yz}}$	\\
	$V_{z0} \cdot \tau^z$	&	scalar potential ($2\times$AII)	\\
	$V_{00} \cdot \openone$	&	scalar potential ($2\times$AII)	&	$\mu = -\xp{V_{00}}$
\end{tabular}
\caption{%
	List of time-reversal invariant disorder terms on the surface of a WTI with two Dirac cones.
	If only one of the disorder structures is present in the system, the type indicates the disorder class of the system and the effect of the disorder.
	For example, with \emph{only} $V_{yz}(\vr)\tau^y\sigma^z$, the system breaks up into two systems, each identical to a Dirac cone with random mass in class D.
	Hence, multiple disorder structures are required for the system to be class AII.
}
\label{tab:DisorderStruct}
\end{table}

\textit{Hamiltonian and disorder structure.}
In the following, we specialize to the case of a WTI with two Dirac cones, for which the low energy electronic properties are described by the
Hamiltonian%
	~\footnote{We choose to work with a Hamiltonian where the two Dirac fermions have the same chirality.
		All the results in our Letter also hold in the case when they have the opposite chirality, due to a similarity transformation between the two cases (\textit{cf.}\ \hyperref[sec:AppB-Chirality]{App.~B}).}
\begin{align}
	H = \hbar v_D \tau^0 (\sigma^x k_x + \sigma^y k_y) + V(\vr),
		\label{eq:H}
\end{align}
where $\tau^0 = \sigma^0 = \openone$ is the identity, $\tau^{x,y,z}$ and $\sigma^{x,y,z}$ are the Pauli matrices in valley and spin space, respectively.
$H$ is invariant under the time-reversal $\Theta = i\sigma^y\mathcal{K}$, where $\mathcal{K}$ is the complex conjugation operator.
The Dirac velocity $v_D$ (taken isotropic for simplicity) and $\hbar$ are set to 1 henceforth.
The disorder potential is written
\begin{equation}
   V(\vr) = \sum_{\alpha\beta} V_{\alpha\beta}(\vr) \, \tau^\alpha \otimes \sigma^\beta  
\end{equation}
with $V_{\alpha\beta}(\vr)$ a scalar potential and $\alpha,\beta \in \{0,x,y,z\}$.
The six terms respecting time-reversal, listed
in Table~\ref{tab:DisorderStruct}, are independently distributed with correlation
\begin{equation}
  \big\langle \delta V_{\alpha\beta}(\vr) \, \delta V_{\alpha\beta}(\vr') \big\rangle = g_{\alpha\beta}K(\vr-\vr')
  \label{eq:g}
\end{equation}
where $\int\! d^2\vr \, K(\vr) = 1$.
The two-terminal conductivity $\sigma$ of a system of size $L$ is obtained numerically by adapting the transfer matrix method of Ref.~\onlinecite{BardarsonSingleParaScaling07} to the current problem.
(The width $W$ is taken large enough that the conductivity is independent of the ratio $W/L$.)
Each disorder term is Gaussian correlated with $K(\vr) = \exp(-r^2/2\xi^2)/(2\pi\xi^2)$.
We also take the averages $\langle V_{\alpha\beta}(\vr) \rangle = 0$,
except for $V_{yz}$ and $V_{00}$ as explained below.

It is useful to first analyze the system in the clean case, where $V_{\alpha\beta}$ are constants.
$\xp{V_{00}}$ acts as the chemical potential $\mu$ which shifts the energy spectrum trivially.
$\tau^y\sigma^z$ anticommutes with all the other potentials (except $\openone$) as well as the kinetic term $\bm\sigma\cdot\vk$; the presence of this term
always gaps the system, and hence we refer to $m = \xp{V_{yz}}$ as the ``mass.''
The energy spectrum of the system is given by
\begin{multline}
	\big( E(\vk) - \mu \big)^2 = k^2 + V_{x0}^2 + V_{yx}^2 + V_{yy}^2 + V_{z0}^2
		\\	\pm 2 \sqrt{ (V_{x0}^2+V_{z0}^2)k^2 + (V_{yx}k_x+V_{yy}k_y)^2 } + m^2 ,
\end{multline}
with minima at $k^2 = V_{x0}^2 + V_{yx}^2 + V_{yy}^2 + V_{z0}^2$ and $k_x/k_y = V_{yx}/V_{yy}$, in which case we have $(E-\mu)^2 = m^2$.
Therefore, the energy gap is $2|m|$ and the system is insulating when $|m| > |\mu| $.
The cases $m > |\mu|$ and $m < -|\mu|$ correspond to the two topological sectors in the 2D AII class, \textit{i.e.}, the trivial and QSH insulator%
	~\footnote{We avoid stating which state $\pm m$ is the trivial insulator and which is the topological (QSH) insulator,
	as this depends on the band structure away from the Dirac point.}.
The intermediate metallic region $-|\mu| < m < |\mu|$ separates the two phases.


In the presence of disorder, a similar description applies --
by varying $m$, one can take the system between the two insulating phases.
As conjugation by $\tau^x$ flips the sign of $m$, a conducting state should be realized at $m=0$.
Because to the stability of the symplectic metal, one does not expect generically a direct transition between the insulating phases%
	~\cite{ObuseQSHNetworkModel07,EssinChernParity07,ShindouMurakami3DQSHDisorder09}.
The resulting phase diagram is shown in Fig.~\ref{fig:QSHMetalIns}.
The shape of the phase diagram around the clean Dirac point $g = m = \mu = 0$ is consistent with the renormalization group flow of the coupling parameters
	$g_{\alpha\beta}$, $m$, and $\mu$ away from that point~(\textit{cf}\ \hyperref[sec:AppA-FlowEq]{App.~A}).
At a finite chemical potential, there is a range
  of mass values $|m| \lesssim |\mu|$ where the system undergoes two transitions with increasing disorder strength (dashed line in
  Fig.~\ref{fig:QSHMetalIns}).
  A similar mass term can be defined for an arbitrary even number of Dirac cones; thus, the phase diagram in Fig.~\ref{fig:QSHMetalIns} holds generally for WTI's (\textit{cf}.\ \hyperref[sec:AppD-EvenDiracCones]{App.~D}).

\begin{figure}[tb!]
	\includegraphics[width=0.96\columnwidth]{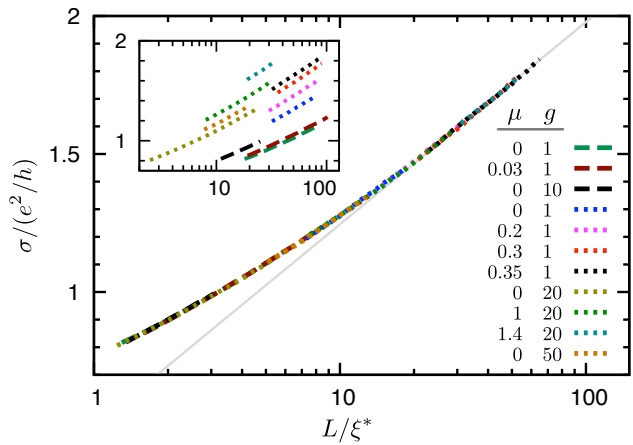}
	\caption{%
		Demonstration of one-parameter scaling at $m = 0$.
		Conductivity as a function of system size for various parameters all collapsed (by shifting the raw data horizontally) onto one scaling curve.
		At large $\sigma$, the slope $d\sigma/d\ln L$ approaches $1/\pi$ (gray line) consistent with weak antilocalization.
		Here $g_{00} = g$ for dotted lines and $g_{00} = 0$ for dashed lines.
			For all other $\alpha\beta$, $g_{\alpha\beta} = g$.
	(Inset)	Raw data $\sigma$ vs.\ $L/\xi$.
	}
	\label{fig:singleparascaling}
\end{figure}

\begin{figure}[tb!]
	\begin{minipage}{\columnwidth}
		\begin{minipage}[c]{5mm}
			\vspace{-46mm}
			\subfigure[]{ \label{fig:sigma-m} }
		\end{minipage}
			\hspace{-7mm}
		\begin{minipage}[c]{\columnwidth}
			\includegraphics[width=0.8\columnwidth]{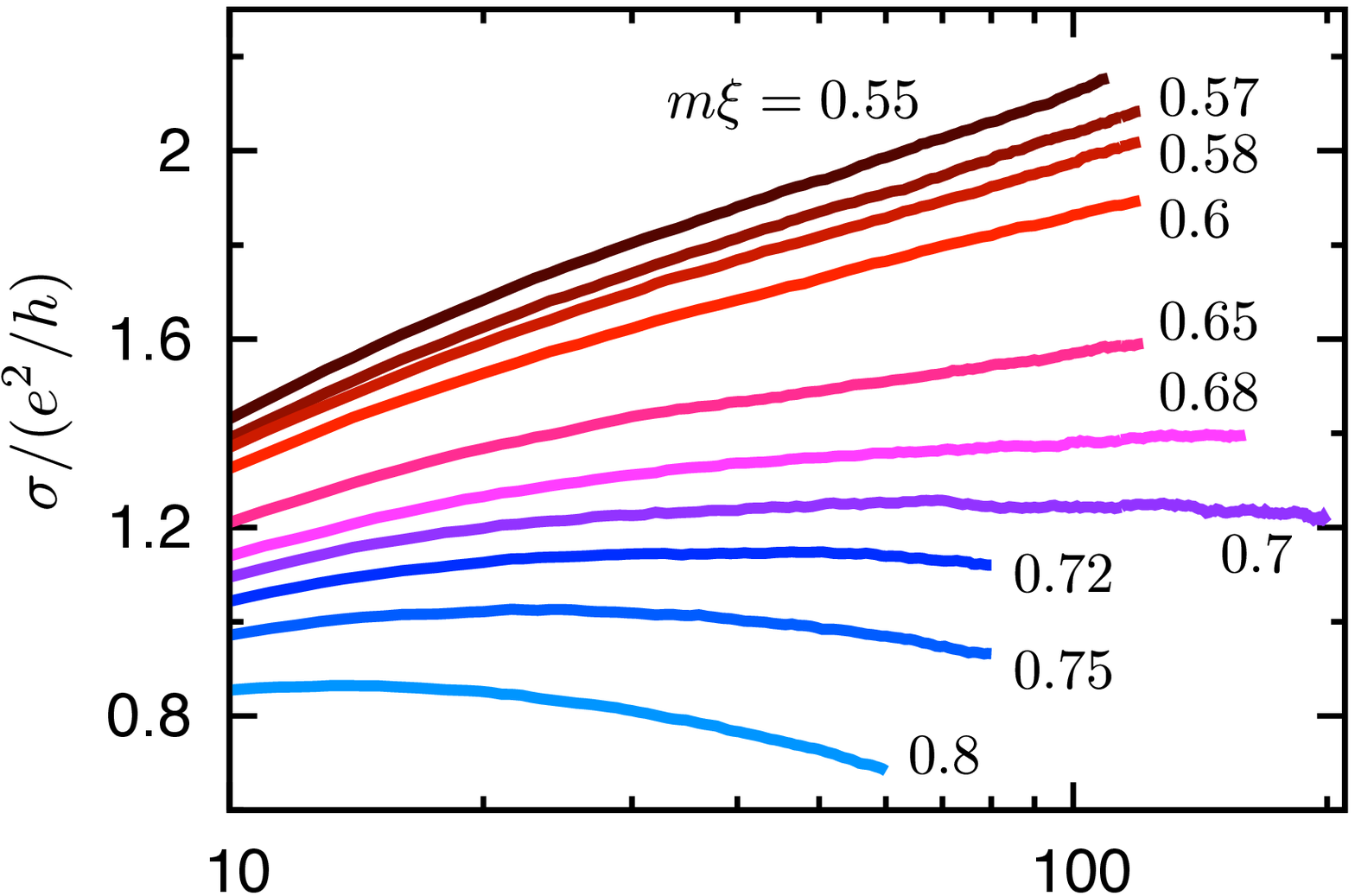}
		\end{minipage}
	\\ \vspace{3mm}
		\begin{minipage}[c]{5mm}
			\vspace{-46mm}
			\subfigure[]{ \label{fig:sigma-K} }
		\end{minipage}
			\hspace{-7mm}
		\begin{minipage}[c]{\columnwidth}
			\includegraphics[width=0.8\columnwidth]{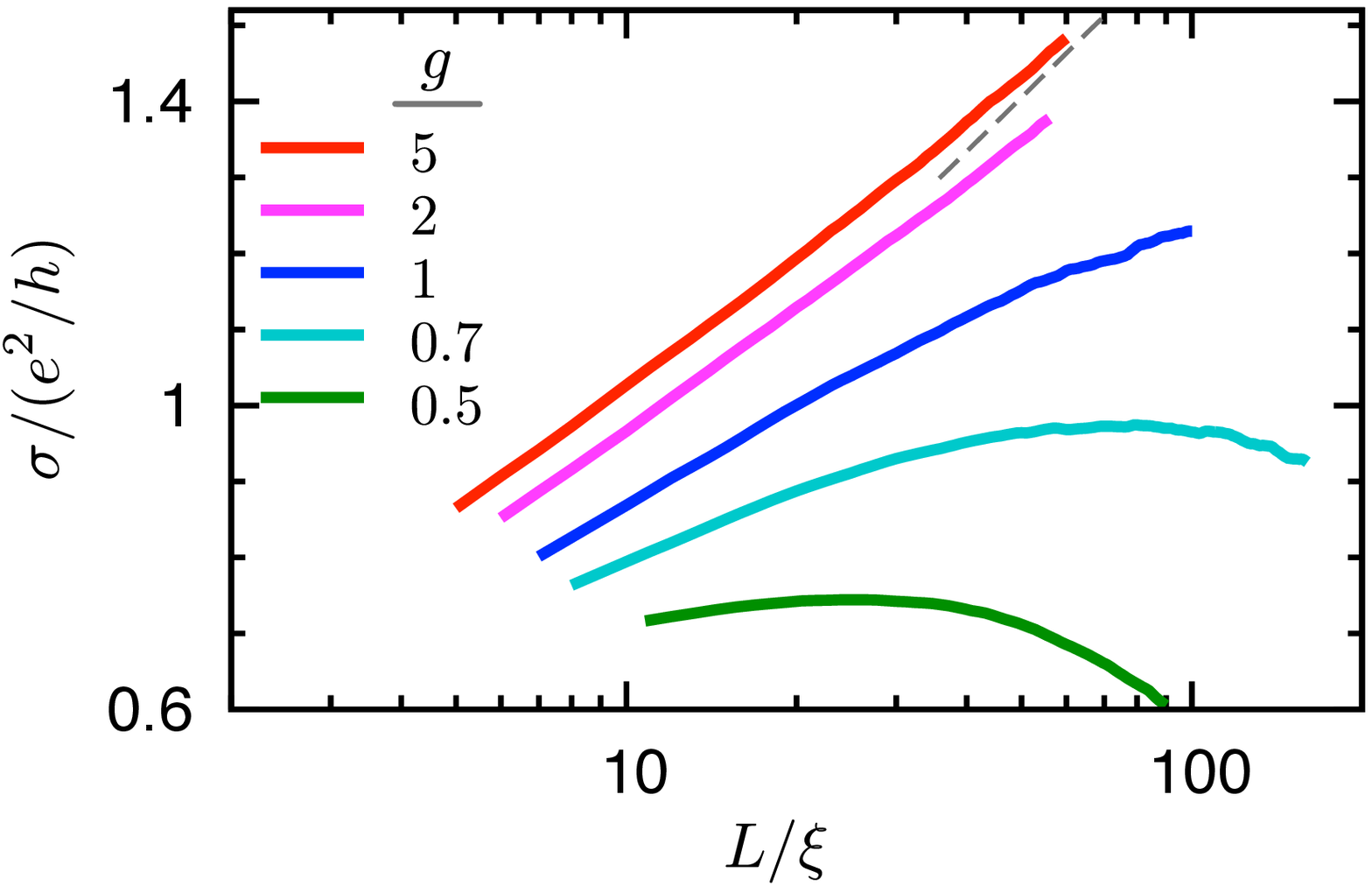}
		\end{minipage}
	\end{minipage}
	\caption{%
	(a)	Metal-insulator transition as $m$ is varied.
		Conductivity is plotted vs.\ system size for fixed $\mu\xi = 1$ and $g_{x0} = g_{yz} = 2$.
		For large $m$ the system flows to an insulating state, while for small $m$ the system is conducting.
		Among the conducting curves, the slope $d\sigma/d\ln L$ approaches $1/\pi$ at large $\sigma$.
		The data show a metal-insulator transition at $\sigma_c$ consistent with the known value of $1.42\,e^2/h$~\cite{MarkosSchweitzerCriticalCond06}.
	(b)	Metal-insulator transition as disorder strength $g_{\alpha\beta} = g$ is varied.
		The plot is $\sigma$ vs.\ $L/\xi$ for fixed $m\xi = 0.05$ and $\mu = 0$.
		Increasing disorder increases the conductivity, inducing a transition from an insulating phase to a metallic one at some critical $g$.
		The dashed line indicates a slope of $1/\pi$.
		These figures are consistent with the phase diagram in Fig.~\ref{fig:QSHMetalIns}.
	}
	\label{fig:sigma-MetalIns}
\end{figure}

The numerical data supporting the phase diagram in Fig.~\ref{fig:QSHMetalIns} are shown in
Figs.~\ref{fig:singleparascaling} and~\ref{fig:sigma-MetalIns}.
At $m=0$ the
conductivity always flows to the symplectic metal, regardless of the strength of the disorder (\textit{cf.}\ Fig.~\ref{fig:singleparascaling}).
By rescaling the length ($L/\xi \rightarrow L/\xi^\ast$), we
can collapse all the data on a single curve demonstrating one-parameter scaling along the $m=0$ line.
At large conductivity, the beta function
$\beta(\sigma) = d(\ln\sigma)/d(\ln L)$
approaches $1/\pi\sigma$ as predicted for weak antilocalization%
	~\cite{Hikami2DDisorder80}. 

By varying $m$, it is possible to drive the system to an insulator, as shown in Fig.~\ref{fig:sigma-m}.
At small $m$, the system remains a symplectic metal.  At some critical $m$, a metal-insulator transition occurs and it ceases to conduct.
For a fixed nonzero $m$ such that the clean system is insulating, disorder drives the system into a metallic phase at some critical
disorder strength $g_c$ that depends on $m$, as demonstrated in Fig.~\ref{fig:sigma-K}.
In both these cases, at large conductivity the slope $d\sigma/d\ln L$ approaches $1/\pi$, indicative of weak antilocalization.

\textit{Conditions for localization.}
Since a WTI is always conducting in the absence of mass, it is pertinent to discuss under what circumstances one expects a nonzero mass. 
The potential $V_{yz}(\vr)$ couples valleys centered at different momenta and thus
requires short-range scatters.
Furthermore a nonzero mass can arise only when the surface potential is commensurate with an even number of unit cells,
such as in the case of cleaving the surface at a crystal plane~\cite{RingelKrausSternWTI11}, or when the WTI is grown on a lattice-matching substrate.
As such, a nonzero mass would be marked by an enlargement of the unit cell and would appear in a crystal diffraction experiment as a peak of order $\vG_\nu/2$,
where $\vG_\nu$ is a reciprocal lattice vector characterizing the weak topological invariants of the WTI%
	~\cite{FuKaneMeleTI3D}.
On the other hand, a period-doubling perturbation could indicate a valley-mixing term other than $m$
(the other possible terms being $\xp{V_{x0}}$, $\xp{V_{yx}}$ or $\xp{V_{yy}}$).
It may also be possible to measure $m$ via spin- and angle-resolved photoemission spectroscopy (spin-resolved ARPES), by comparing the spin-up and spin-down intensities at wavevector $\vG_\nu/2$.
This proposal is motivated by the form of the potential $\tau^y\sigma^z$, which differentiates the up and down spins.
Localization may also occur due to lattice effects or higher order terms in the Hamiltonian~\cite{ObuseQSHNetworkModel07,YamakageTIMultipleDisorderTransitions11,RyuHighGradientOperatorsWZW10}.

In the case where the WTI consists of an odd number of QSH layers, we argue that the mass must be identically zero.
Consider stacking $n$ QSH layers, with each layer in the $\va_1, \va_2$ plane, and the layers $\va_3$ offset from one another.
For simplicity, we impose a periodic boundary condition in the $\va_3$ direction.
The surface spectrum of a plane parallel to $\va_3$ will have two Dirac cones, centered on different time-reversal invariant momenta $\vk_a$ and $\vk_b$, such that $(\vk_b - \vk_a) \cdot \va_3 = (\vG_\nu/2) \cdot \va_3 = \pi$.
The second quantized kinetic Hamiltonian will be of the form
$\Psi_a^\dag (\vk-\vk_a)\cdot\bm\sigma \Psi_a + \Psi_b^\dag (\vk-\vk_b)\cdot\bm\sigma \Psi_b$.
$\Psi^\dag$ and $\Psi$ are the creation and annihilation operators satisfying the boundary condition $\Psi(\vr + n\va_3) = \Psi(\vr)$.
To cast this into the form of the effective Hamiltonian \eqref{eq:H}, we perform the gauge transformation $\Psi_\mu \rightarrow \Psi_\mu e^{i \vk_\mu \cdot \vr}$ for each of the fermion species.
The gauge transformation will, in general, change the boundary condition for the operators $\Psi_a$ and $\Psi_b$.
Notice that $\exp[ i (\vk_a - \vk_b) \cdot (n\va_3) ] = (-1)^n$, and hence for odd $n$ the transformed operators will have differing boundary conditions: \textit{i.e.}, one periodic and one antiperiodic.
The mass term coupling the fermion species together in the effective Hamiltonian must have antiperiodic boundary conditions and, hence, averages to zero.
Therefore, for an odd number of stacked QSH layers, $m$ is zero and the surface (parallel to the stacking direction) always flows to a metallic phase.
These results settle the question of which of the two
possible flow diagrams consistent with the quasi-1D numerics in Ref.~\onlinecite{RingelKrausSternWTI11} is actually realized.

\begin{figure}[tb!]
	\begin{minipage}{\columnwidth}
		\begin{minipage}[c]{5mm}
			\vspace{-40mm}\hspace{-5mm} \subfigure[]{ \label{fig:TwoParaFlow} }
		\end{minipage}
			\hspace{-6mm}
		\begin{minipage}[c]{0.35\columnwidth}
			\includegraphics[width=\columnwidth]{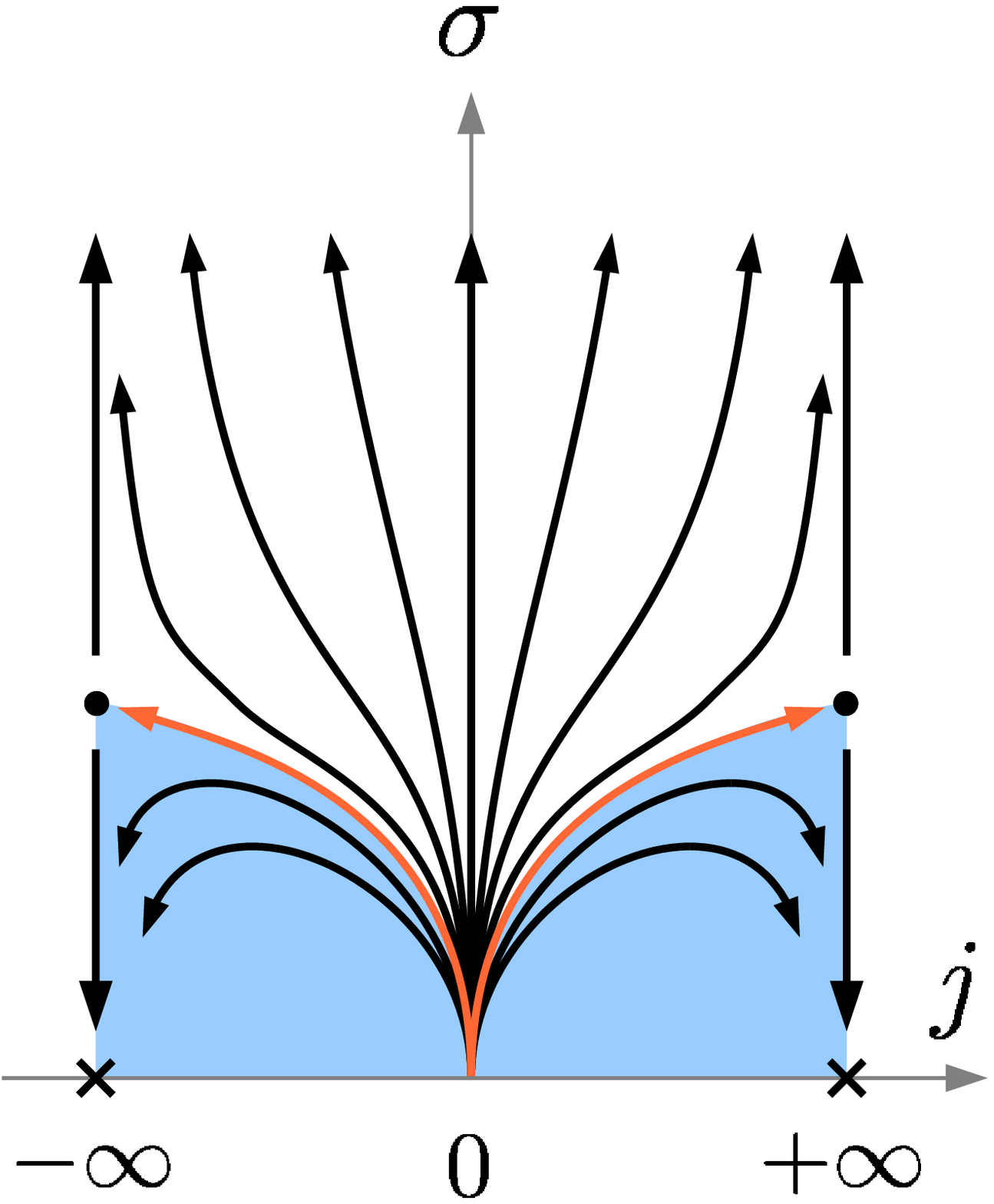}
		\end{minipage}
		\begin{minipage}[c]{5mm}
			\vspace{-40mm} \subfigure[]{ \label{fig:TwoParaData} }
		\end{minipage}
			\hspace{-7mm}
		\begin{minipage}[c]{0.58\columnwidth}
			\includegraphics[width=\columnwidth]{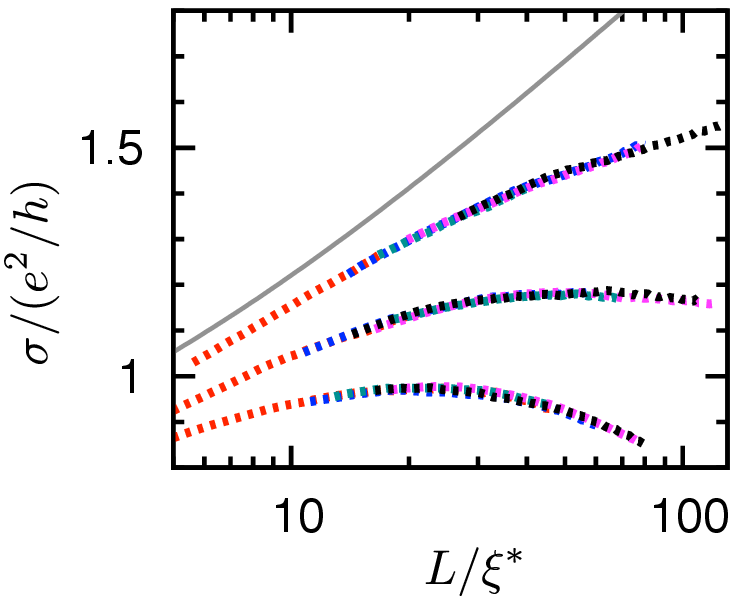}
		\end{minipage}
	\end{minipage}
	\caption{%
	(a)	Two-parameter flow which captures the QSH-metal-insulator transition in the AII class.
		The scaling variables are $\sigma$ and $j$, the latter of which separates the normal or QSH insulator phases.
	(b)	Numerical data $\sigma(L/\xi^\ast)$, demonstrating that the conductivity curves may be collapsed onto each other.
		The gray line is the data at $m=0$ from Fig.~\ref{fig:singleparascaling}.
		The raw data and parameters are given in \hyperref[sec:AppE-2ParaRawData]{App.~E}.
	}
	\label{fig:TwoPara}
\end{figure}

\textit{Two-parameter scaling.}
The existence of one-parameter scaling along the line $m = 0$ suggests that there might be a two-parameter scaling collapse for the entire range of parameters when the mass is nonzero, analogous to the quantum Hall transition (in the A class)%
	~\cite{KhmelnitskiiQHConductivity83,PruiskenQHThetaTerm84,PruiskenDiagram85}.
Figure~\ref{fig:TwoParaFlow} shows a possible flow for conductivity $\sigma$ and the (unknown) second scaling parameter $j$.
The horizontal scale $j$ distinguishes between the two topological phases, much in the same way as the Hall conductivity in the quantum Hall case.

Even without a precise definition of $j$ as an experimental quantity, we may still infer a number of properties of the flow diagram.
(i) For large conductivity $\sigma$, $\beta(\sigma)$ is positive and $\sigma$ flows upward towards infinity.
(ii) There are two insulating stable fixed points (crosses) at $(\sigma,j) = (0,\pm\infty)$ and regions which flows toward them (shaded regions).
(iii) Consequently, there must be unstable fixed points (dots) at $j = \pm\infty$ which mark a metal-insulator transition.
(iv) Near $j = 0$, the system must flow to a metallic phase, as there should not be a direct phase transition between the two insulating phases.
Figure~\ref{fig:TwoParaFlow} gives the simplest flow diagram consistent with these requirements.

The two-parameter scaling of $(\sigma, j)$ implies that $\sigma(L/\xi)$ cannot be collapsed onto a single scaling curve (as in Fig.~\ref{fig:singleparascaling}), but onto a family of curves parameterized by a single variable $x$.
The scaling form is
\begin{align}
	\sigma = f(L/\xi^\ast; x),
	\label{eq:2parascalingform}
\end{align}
where all the microscopic parameters $m$, $\mu$, $g_{\alpha\beta}$, $\xi$, \textit{etc}., determine the conductivity only via the two functions $x$ and $\xi^\ast$.

In Fig.~\ref{fig:TwoParaData}, we present the accompanied data for our two-parameter scaling hypothesis, by collapsing $\sigma$ vs.\ $L/\xi^\ast$ onto a family of curves.
For each curve, the parameters $\mu$ and $g_{\alpha\beta}$ were fixed while $m$ is varied until $\sigma(L)$ overlaps with the existing set of curves.
The data show reasonable agreement with the scaling form \eqref{eq:2parascalingform}.

Quantum transport at the surface of a weak topological insulator thus shows a scaling structure similar to that of the quantum Hall plateau transitions.
It should be possible to interpret experiments on weak topological insulators in terms of the above Dirac model and possibly to control the parameter $m$
by choosing a substrate whose lattice potential generates the massive perturbation.
In addition, the electronic structure of thin films of STI's can be
mapped to the two Dirac cone system studied here, with the tunneling between the surfaces taking the role of the mass (\textit{cf}\ \hyperref[sec:AppC-STIThinFilm]{App.~C}).
Our results should motivate the
search for WTI candidate materials, of which there are few.
It remains to be seen if the two-parameter flow is generic to all noninteracting disordered systems in the symplectic class.

We gratefully acknowledge insightful discussions with Christopher Mudry, Shinsei Ryu, and Ashvin Vishwanath.
This work is supported by FENA (R.M.), DOE BES (J.H.B.), and NSF DMR-0804413 (J.E.M.).


\appendix
\clearpage

\onecolumngrid

\section{A. Renormalization flow away from clean Dirac point}
\label{sec:AppA-FlowEq}
\noindent
Disorder average, either using the replica trick or a supersymmetric representation, gives arise to an interacting field theory with coupling constants
given by the amplitudes $g_{\alpha\beta}$ of the disorder correlator, as well as single particle potentials with amplitudes $\mu$ and $m$%
	~\cite{EfetovSSDisorderChaos,AltlandSimons2nd}.
The clean Dirac point $g_{\alpha\beta} = m = \mu = 0$ is a fixed
point of this theory. Under renormalization the coupling constants flow away from the clean Dirac point according to the equations: 
\begin{align}
\begin{split}
	\pi\frac{dg_{00}}{d\ln L} &= g_{00}(g_{00} + g_{x0} + g_{yx} + g_{yy} + g_{yz} + g_{z0}) + g_{yz}(g_{yx} + g_{yy})	, \\
	\pi\frac{dg_{x0}}{d\ln L} &= g_{x0}(g_{00} + g_{x0} - g_{yx} - g_{yy} - g_{yz} - g_{z0}) + g_{z0}(g_{yx} + g_{yy})	, \\
	\pi\frac{dg_{yx}}{d\ln L} &= g_{x0}g_{z0} + g_{yz}g_{00}
		= \pi\frac{dg_{yy}}{d\ln L}	, \\
	\pi\frac{dg_{yz}}{d\ln L} &= g_{yz}(-g_{00} + g_{x0} + g_{yx} + g_{yy} - g_{yz} + g_{z0}) + g_{00}(g_{yx} + g_{yy})	, \\
	\pi\frac{dg_{z0}}{d\ln L} &= g_{z0}(g_{00} - g_{x0} - g_{yx} - g_{yy} - g_{yz} + g_{z0}) + g_{x0}(g_{yx} + g_{yy})	, \\
	\frac{d\mu}{d\ln L} &= \mu + \frac{\mu}{2\pi}(g_{00} + g_{x0} + g_{yx} + g_{yy} + g_{yz} + g_{z0})	, \\
	\frac{dm}{d\ln L} &= m + \frac{m}{2\pi}( - g_{00} + g_{x0} + g_{yx} + g_{yy} - g_{yz} + g_{z0})	.
\end{split}
\label{eq:RG}
\end{align}
These equations have been obtained in one loop following the standard procedure used for Dirac fermions~%
	\cite{CardyRenormalization,AltlandSimonsZirnbauerdwaveSC02,OstrovskyMirlinGphDisorder06,SchuesslerMirlinGphBallistic09}.
Since $m$ initially flows away faster than
the disorder couplings, this suggest the shape of the phase diagram around the clean point is as in Fig.~2 in the main text.
Other interesting features
of these equations are that both the random scalar potential $g_{00}$ and random mass $g_{yz}$ tend to decrease the mass, and thereby enhance the
conductivity.
In contrast, all the other disorder structures tend to increase the mass. 

Apart from these observation and that the clean fixed point is unstable, the renormalization group equations~\eqref{eq:RG} yield no more information on the metal-insulator transition.
Therefore the numerical simulations reported on in the main text are crucial in order to answer such questions.

\section{B. Chirality of the Dirac cones}
\label{sec:AppB-Chirality}
\noindent
In the main text we used a model in which the two Dirac cones have the same chirality:
\begin{align}
	H_0 = \hbar v_D \begin{bmatrix} \sigma^x k_x + \sigma^y k_y & \\ & \sigma^x k_x + \sigma^y k_y \end{bmatrix} .
\end{align}
Here $H_0$ refers to the kinetic portion of the Hamiltonian.
As the momentum $\vk$ rotates by $2\pi$, the spin also rotates by $2\pi$ in the same direction, hence both Dirac cones have chirality of $+1$.

We can apply the unitary transformation $U$ which flips only one of the Dirac cone's chirality.
\begin{align}
	U &= \begin{bmatrix} \openone & \\ & i\sigma^x \end{bmatrix}
	&	\Rightarrow\quad	H'_0 = U H_0 U^\dag
		&	= \hbar v_D \begin{bmatrix} \sigma^x k_x + \sigma^y k_y & \\ & \sigma^x k_x - \sigma^y k_y \end{bmatrix} .
\end{align}
It is important to note that the transformation does not alter the form of time-reversal, \textit{i.e.}~$U \Theta U^\dag = \Theta = i\sigma^y\mathcal{K}$,
and that it shuffles the disorder potentials:
\begin{align}
	(V_{x0}, V_{yx}, V_{yy}, V_{yz}, V_{z0}, V_{00})
	\rightarrow (V_{yx}, -V_{x0}, -V_{yz}, V_{yy}, V_{z0}, V_{00}) .
\end{align}
The mass for the new system is defined to be $m = \xp{V_{yy}}$ as $\tau^y\sigma^y$ anticommutes with $H'_0$.
This shows that our results are independent of the chirality of the Dirac cones.

\section{C. Mapping to strong topological insulators thin films}
\label{sec:AppC-STIThinFilm}
\noindent
A thin film of strong topological insulator consists of two Dirac cones, for which the two cones will couple to each other for a sufficiently thin sample.
The low-energy Hamiltonian for the thin film (with inversion symmetry) \cite{CXLiuOscTI2D3DCrossover10,GhaemiTopothermo} is
\begin{align}
	H_\textrm{STI} = \begin{bmatrix} \hbar v_D \bm\sigma \times \vk & \Delta
				\\ \Delta & -\hbar v_D \bm\sigma \times \vk \end{bmatrix} + V(\vr) ,
\end{align}
where $\bm\sigma \times \vk = \sigma^x p_y - \sigma^y p_x$, and $\Delta$ is the tunneling amplitude between the two surfaces.
The Hamiltonian is written in the basis $(\textrm{t}\!\uparrow, \textrm{t}\!\downarrow, \textrm{b}\!\uparrow, \textrm{b}\!\downarrow)$, where $\textrm{t}$ and $\textrm{b}$ represents the excitations for the top and bottom surfaces respectively.

The model for WTI may be mapped to $H_\textrm{STI}$ via the transformation
\begin{align}
	H_\textrm{STI} & = U H U^\dag ,
		& \textrm{where}\;
			U & = \exp \left[ i\tfrac{\pi}{4} \tau^z \sigma^z \right] .
\end{align}
(Again note that $U \Theta U^\dag = \Theta = i\sigma^y\mathcal{K}$.)
The mass corresponding to this system is $m = \xp{V_{x0}} = \Delta /\hbar v_D$, which is a measure of the tunneling amplitude between the surfaces.
Consequently, the physics described in the main text may also be realized experimentally by a thin film of strong topological insulator, where the film's thickness can be used to tune $m$.
(The mass $m$ may oscillate between being positive and negative as the thickness is varied, as shown in Ref.~\onlinecite{CXLiuOscTI2D3DCrossover10}.)

\section{D. Constructing the mass for an arbitrary even number of Dirac cones}
\label{sec:AppD-EvenDiracCones}
\noindent
For the case of two Dirac cones, we have defined the mass $m = \xp{V_{yz}}$ with the following properties:
(\textbf{1}) In absence of other potentials and disorder, the system gap is simply $2|m|$, and
(\textbf{2}) the limits $m \rightarrow \infty$ and $m \rightarrow -\infty$ correspond to the system being in the trivial and QSH insulating phases, respectively.
Here we explicitly construct $m$ for a system with four Dirac cones and give the procedure for finding $m$ in the general case.

With four Dirac cones, the Hamiltonian is
\begin{align}
	H^{(4)} = \upsilon^0\tau^0 \bm\sigma\cdot\vk + V(\vr).
\end{align}
We use the Pauli matrices $\tau^{x,y,z}$ and $\upsilon^{x,y,z}$ acting in valley space to span all the possible intervalley couplings.
($\tau^0 = \upsilon^0 = \openone$.)
The disorder potential decomposes
\begin{align}
	V(\vr) & = \sum_{\alpha\beta\gamma} V_{\alpha\beta\gamma}(\vr) \upsilon^\alpha \tau^\beta \sigma^\gamma ,
		\quad \textrm{with } \alpha,\beta,\gamma \in \{0,x,y,z\}.
\end{align}
Of the twenty-eight disorder structures $\upsilon^\alpha \tau^\beta \sigma^\gamma$ compatible with time-reversal $\Theta = i\sigma^y\mathcal{K}$, only six anticommutes with the kinetic Hamiltonian $\bm\sigma\cdot\vk$.
These six are given by $(\alpha,\beta,\gamma) = (0,y,z),(y,x,z),(y,z,z),(y,0,z),(x,y,z),(z,y,z)$.
The first three anticommute with each other as do the last three, while all of the first three commute with any of the last three.
The mass is thus given by:
\begin{align}
	m^{(4)} = \sqrt{\xp{V_{0yz}}^2 + \xp{V_{yxz}}^2 + \xp{V_{yzz}}^2}
					- \sqrt{\xp{V_{y0z}}^2 + \xp{V_{xyz}}^2 + \xp{V_{zyz}}^2} .
\end{align}
It can be shown that this definition satisfies (\textbf{1}) and (\textbf{2}) above.

In the general case with $2n$ Dirac cones, there will be $n(2n-1)$ linear independent disorder structures which opens a gap in the system, of the form $V \otimes \sigma^z$, where $V$ is an $2n \times 2n$ matrix acting in valley space.
$V$ must be an antisymmetric pure-imaginary matrix, \textit{i.e.}~an element of the Lie algebra $\mathfrak{so}(2n)$ (in the canonical representation).
Seeing that $V\otimes\sigma^z$ anticommutes with $\bm\sigma\cdot\vk$, the spectrum at $\vk=0$ is the set of eigenvalues of $V$.
The mass is defined as follows: $|m|$ is the smallest non-negative eigenvalue of $V$ while the sign of $m$ is that of $\operatorname{Pf}(iV)$.

\section{E. Raw data for two-parameter scaling}
\label{sec:AppE-2ParaRawData}
\noindent
\begin{minipage}{\columnwidth}
\center{
	\includegraphics[height=0.3\columnwidth]{twoparascaling}
	\hspace{2mm}
	\includegraphics[height=0.3\columnwidth]{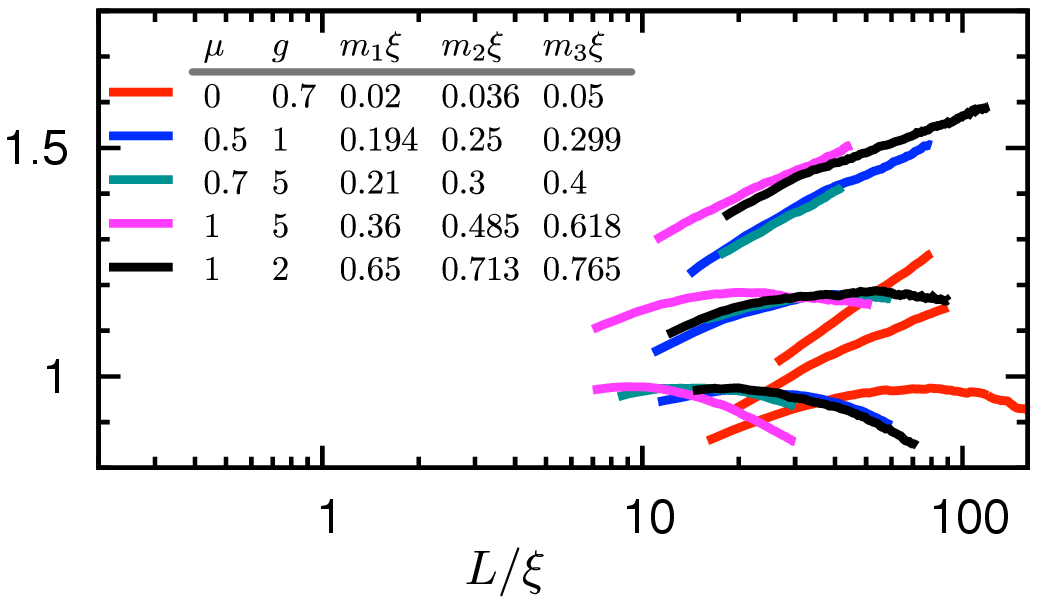}
}
\end{minipage}
\begin{minipage}{\columnwidth}
\vspace{2mm}
\small{
	(Left)	Numerical data $\sigma(L/\xi^\ast)$, demonstrating that the conductivity curves may be collapsed on to each other.
		The gray line is the data at $m=0$ from Fig.~2 of the main text.
	(Right) Raw data $\sigma(L/\xi)$.
		For each data set (denoted by color), the parameters $\xi$, $\mu$, $g_{\alpha\beta}$ are fixed.
		The black data set has $g_{x0} = g_{yz} = g$, while for other colors, $g_{\alpha\beta} = g$ for all $\alpha\beta$.
		Within each data set, $m$ is varied until $\sigma(L/\xi^\ast)$ fit on top of one another.
		The numerical values are given as $m_1$, $m_2$, $m_3$.
		(Increasing $m$ decreases the conductivity, hence the top curve of a data set has the smallest $m$.)
	}
\end{minipage}

\section{F. Conductance Distribution}
\label{sec:AppF-CondDist}
\noindent
\begin{minipage}{\columnwidth}
\center{
		\includegraphics[width=170mm]{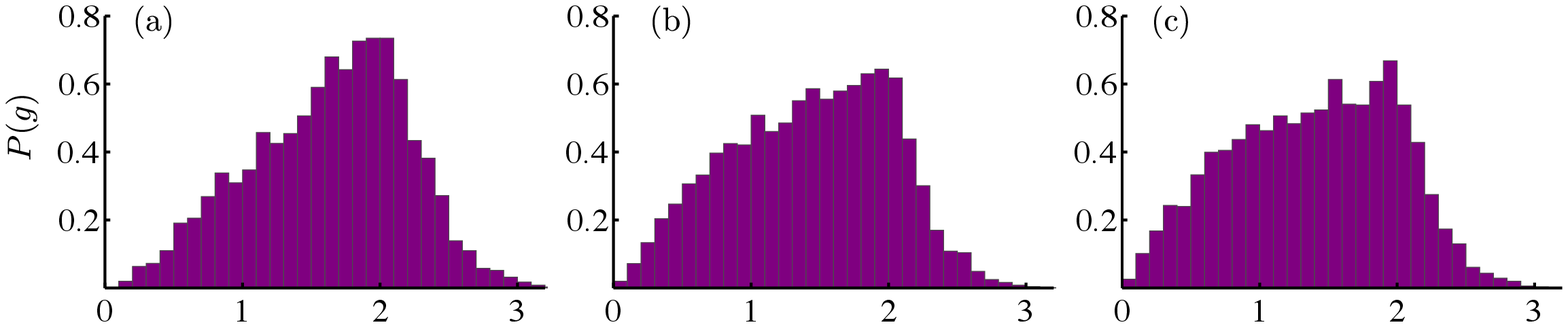}
	\\[2mm]
	\begin{minipage}{119mm}
		\includegraphics[width=115mm]{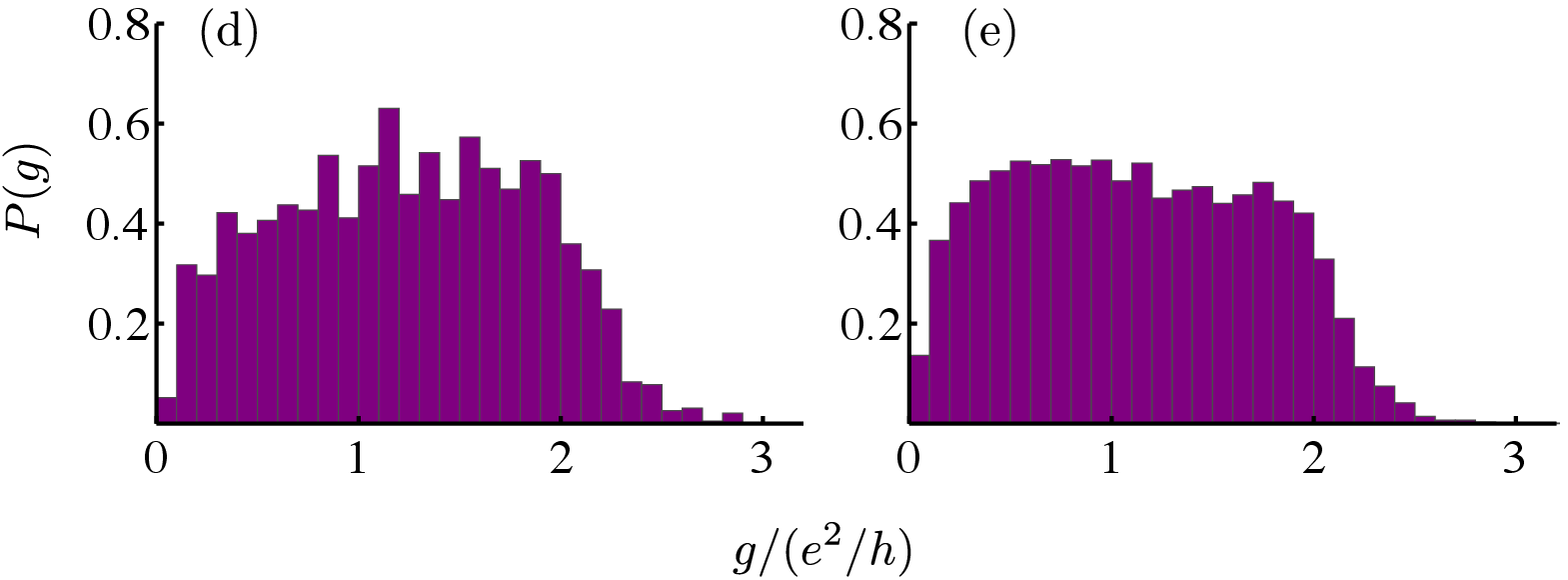}
	\end{minipage}
	\begin{minipage}{50mm}
		(f)\\[1mm]
		\small{
		\begin{tabular}{c |@{\quad} c @{\quad} c @{\quad} c}
			histogram	&	$m\xi$	&	$L/\xi$	&	samples	\\
		\hline
			(a)	&	0.65	&	160	&	3456	\\
			(b)	&	0.67	&	80	&	9600	\\
			(c)	&	0.68	&	160	&	3456	\\
			(d)	&	0.7	&	160	&	1920	\\
			(e)	&	0.72	&	80	&	9600	\\
		\end{tabular}
		}
		\\[2mm] $\phd\quad \mu\xi = 1$, $g_{x0} = g_{yz} = 2$.
		\\ \phd
	\end{minipage}
}
\end{minipage}

\begin{minipage}{\columnwidth}
\vspace{2mm}
\small{
(a)-(e) The conductance distribution $P$ as a function of the conductance $g$ for a square geometry ($W=L$).
The parameters used are those of Fig.~\ref{fig:sigma-m} with varing $m$, given in the table (f).
Comparing with the critical conductance distribution in
	Ref.~\onlinecite{OhtsukiAndersonTransCondDist04} computed for network models,
	we see that the metal-insulator transition occurs between (b) and (c) ($0.67 < m\xi < 0.68$).
	This confirms that the 2D Anderson transition in the symplectic class is universal among both network models and continuous models with smooth disorder.
}
\end{minipage}



\bibliography{weakTI}

\end{document}